\documentclass[preprint]{revtex4}
\usepackage{enumerate}
\usepackage{graphicx}
\usepackage{dcolumn}
\usepackage{bm}
\usepackage{amsmath}
\usepackage{amstext}
\usepackage{amssymb}
\usepackage{latexsym}
\usepackage{amsfonts}
\usepackage{amsmath}
\usepackage{amssymb}
\usepackage{graphicx}
\providecommand{\U}[1]{\protect\rule{.1in}{.1in}}

\begin{document}
\title{Minimum Dissipation Principle in Nonlinear Transport}
\author{Giorgio Sonnino${}^{1,2\star}$, Jarah Evslin${}^{3\star}$, Alberto Sonnino${}^{4,5\star}$
}
\address{
\noindent ${}^1$Universit{\'e} Libre de Bruxelles (U.L.B.), Department of Theoretical Physics and Mathematics, Campus Plaine C.P. 231 Brussels - Belgium\\
\noindent${}^2$Royal Military School (RMS), Av. de la Renaissance 30 1000 Brussels - Belgium\\
\noindent${}^{3}$High Energy Nuclear Physics Group, Institute of Modern Physics, Chinese Academy of Sciences, Lanzhou - China\\
\noindent${}^{4}$Karlsruhe Institute of Technology (KIT), Department of Electrical Engineering and Information Technology (ETIT), Campus Süd Engesserstrae 13 D-76131 Karlsruhe - Germany\\
\noindent${}^{5}$Ecole Polytechnique de Louvain (EPL), Universit{\'e} Catholique de Louvain (UCL), Rue Archim{\`e}de, 1 bte L6.11.01, 1348 Louvain-la-Neuve - Belgium
}
\begin{abstract}
We extend Onsager's minimum dissipation principle to stationary states that are only subject to local equilibrium constraints, even when the transport coefficients depend on the thermodynamic forces.  Crucial to this generalization is a decomposition of the thermodynamic forces into those that are held fixed by the boundary conditions, and the subspace which is orthogonal with respect to the metric defined by the transport coefficients.  We are then able to apply Onsager and Machlup's proof to the second set of forces.  As an example we consider two-dimensional nonlinear diffusion coupled to two reservoirs at different temperatures.  Our extension differs from that of Bertini, \textit{et al.} in that we assume microscopic irreversibility and we allow a nonlinear dependence of the fluxes on the forces.
\\
\\
\\
\\
\\
\noindent ${}^{\star}$ Email: gsonnino@ulb.ac.be, jarah@ihep.ac.cn, alberto.sonnino@gmail.com

\end{abstract}

\maketitle

\section{\bf Introduction. Relaxation to Equilibrium}
Equilibrium statistical thermodynamics is able to estimate the macroscopic quantities and their fluctuations without solving the equation of motion. However, an analysis of dynamical properties of the systems is essential if we are interested in determining the macroscopic behaviour of the systems relaxing to non-equilibrium steady states or in obtaining the probability distributions of fluctuations. In such cases, the Gibbs distribution has to be replaced by a suitable distribution of fluctuations valid for the microscopic dynamics. This theory should be based on the extension of the Boltzmann-Einstein equilibrium fluctuation theory combined with dynamics. One example of such example is the {\it macroscopic fluctuation theory} of Ref.~\cite{bertini}. This theory has been used for studying some microscopic models and it leads to various interesting predictions \cite{appert}-\cite{derrida}. In this paper we develop a theory for non-equilibrium steady states mainly based on the Boltzmann-Einstein theory. Crucial in our approach is the decomposition of the thermodynamic forces that are held fixed by the boundary conditions (fixed thermodynamic forces) into those, which have no external interactions (free thermodynamic forces). We shall show that, without explicit knowledge of the entire invariant distribution function for the microscopic dynamics, the Onsager-Machlup functional, restricted only to the free thermodynamic forces, approximates the probability of a particular relaxation to a stationary state. We also provide an estimation of the error. 

\noindent The Boltzmann-Einstein theory received a rigorous mathematical formulation in classical equilibrium statistical mechanics via the so called large deviation theory (LDT) \cite{lanford}. The LDT has also been applied to hydrodynamic evolutions of stochastic interacting particle systems \cite{kipnis} and extended to nonlinear hydrodynamic regime \cite{jona}. A general theory of large deviations for irreversible processes, i.e. when the detailed balance does not hold, has been successively formulated in 2002 by  Bertini {\it et al.} \cite{HJ1}-\cite{eyink}.  Several examples of LDT are provided by thermodynamic systems driven towards non-equilibrium steady-states by the boundary conditions. This is the case, for example, of a fluid in contact within two thermal reservoirs where a flow of matter, or heat, through the system is established. Bertini {\it et al.} showed that the spontaneous fluctuations of the process is described by the time reversed dynamics \cite{HJ1}. In Ref.~\cite{HJ1} it is shown that the violation of the Onsager-Machlup symmetry observed, for example, in stochastically perturbed reversible electronic devices \cite{luchinsky}, is also connected to the time reversed dynamics. 

\noindent When the thermodynamic forces, indicated with ${X}^\mu$, are nonzero, entropy $s$ is produced according to the balance equation
\begin{equation}\label{j2}
\frac{\partial s}{\partial t} + \nabla\cdot I = \frac{d_Is}{dt}
\end{equation}
where $I$ and $d_Is$ are the reversible entropy flow and the density of entropy production, respectively. Let us consider now a system characterized by $n$ degrees of variables $A_1,\ A_2,\cdots A_n$. The (local) equilibrium values are $A_1^0,\ A_2^0,\cdots A_n^0$. Denoting by $\alpha_\mu=A_\mu-A_\mu^0$ ($i=1\cdots n$) the $n$ deviations of the thermodynamic quantities from their equilibrium value (fluctuations), Prigogine proposed that the probability distribution of finding a state in which the values $\alpha_\mu$ lie between $\alpha_\mu$ and $\alpha_\mu+d\alpha_\mu$ is given, up to a normalization constant, by \cite{prigogine}
\begin{equation}\label{i3a}
{\mathcal F}\ \propto\textup{exp}\Bigl(-\frac{\Delta_I S}{k_B}\Bigr)
\end{equation}
\noindent with $k_B$ denoting Boltzmann's constant. The negative sign in Eq.~(\ref{i3a}) is due to the fact that, during the processes, $-\Delta_IS\leq0$. Indeed, if $-\Delta_IS$ were positive, the transformation $\alpha_i\rightarrow \alpha'_i$ would be a spontaneous irreversible change and thus be incompatible with the assumption that the initial state is a (local) equilibrium state \cite{prigogine}. The Prigogine theory generalizes the Boltzmann-Einstein theory \cite{boltzmann}, \cite{landau}, which applies only to equilibrium thermodynamic fluctuations or to adiabatic transformations \cite{prigogine}, \cite{degroot}, \cite{GP71}. The thermodynamic forces, $X^\mu$, and the conjugate fluxes, $J_\mu$, are related to thermodynamic fluctuations by
\begin{equation}\label{i4a}
{X}^\mu=\frac{\partial \Delta_IS}{\partial \alpha_\mu}\qquad ; \quad J_\mu={\dot\alpha}_\mu
\end{equation}
\noindent where the dot stands for the substantial time derivative. Here we have followed Onsager and Machlup$'$s notation, treating the forces as if they were scalars as in the case of chemical reactions. However the same arguments apply to vector and tensor forces obtained by considering as thermodynamic forces the single components of these quantities. 

\noindent  An equilibrium state is stable if a perturbation in the densities $\alpha_\mu$ leads to a restoring force $X^\mu$. The restoring force creates a current $J_\mu=\dot\alpha_\mu$ which cancels the perturbation. Near equilibrium Onsager \cite{Onsager} has extended a theorem of Lord Rayleigh \cite{Rayleigh} that this process is completely characterized by the minimization of a dissipation functional with respect to a variation of the currents. In this note we will extend Onsager and Machlup's stochastic derivation of this result \cite{OM} to more general processes in which the system is connected to multiple reservoirs at which certain intensive variables are held fixed. These reservoirs prevent the system from arriving at equilibrium, as there will be a flux of the corresponding extensive quantities through the system from one reservoir to another. However the system will nonetheless, under certain conditions, arrive at a stationary state which is locally in equilibrium, and we will extend Onsager and Machlup's argument to demonstrate that this relaxation process is characterized by the minimization of only those parts of the dissipation functional corresponding to an orthogonal subset of the forces to the residual current. To make this story quantitative, one needs to know the forces $X^\mu$ as a function of the extensive variables $\alpha_\mu$ and the currents $J_\mu$ as a function of the forces $X^\mu$.  While the system is no longer in equilibrium, we will assume that locally it is still in equilibrium, and so one may locally define the total entropy density $s$.  In general, the thermodynamic fluxes are functions of the forces. At the thermodynamical equilibrium the thermodynamic forces are zero and the fluxes vanish too. The transport equations ({\it i.e.,} the flux-force relation) may be brought into the form
\begin{equation} \label{J}
J_\mu=g_{\mu\nu}(X)X^\nu
\end{equation}
\noindent where $X^\mu$ denotes the thermodynamic forces per unit volume and the elements of the matrix $g_{\mu\nu}$ are identified with the transport coefficients. In Eqs~(\ref{J}), as well as in the sequel, Einstein's summation convention on the repeated indexes is implicitly understood. Given the function $g_{\mu\nu}$ and a configuration $\alpha_\mu$ one may calculate ${\dot\alpha}_\mu$ and so determine the entire future evolution of the system. When the forces are small, corresponding to small deviations from equilibrium, one may approximate $g_{\mu\nu}$ to be independent of $X^\mu$, corresponding to fluxes $J_\mu$ which are linear in $X^\mu$. We refer to this region as the {\it linear Onsager region}. In this region, Onsager noted that the phenomenological linear relation (\ref{J}), which gives ${\dot\alpha}_\mu$ as a function of $\alpha_\mu$, maximizes the functional
\begin{equation}
M=\int_\Omega \dot s\ d{\bf x}- \frac{1}{2} \int_\Omega g^{\mu\nu}{\dot\alpha}_\mu{\dot\alpha}_\nu d{\bf x}
\end{equation}
varied with respect to ${\dot\alpha}_\mu$ everywhere except for the boundaries (with $d{\bf x}$ denoting a spatial volume element, and the integration is over the entire space $\Omega$ occupied by the system). In a dynamical context, one may ask what is the most probable trajectory followed by the system in the spontaneous emergence of a fluctuation or during the relaxation towards equilibrium. Under the assumption of time reversibility and by using a stochastic argument, Onsager and Machlup, demonstrate that the most probable trajectory is obtained by minimizing the quantity $M$ in the situation of a linear macroscopic equation (i.e., very close to equilibrium) \cite{OM}, \cite{MO}. One may now extend the question by asking, for example, what is the most probable trajectory followed by the system in the relaxation to the boundary driven stationary non-equilibrium states. In their paper, Bertini {\it et Al.} formulated a dynamical fluctuation theory for non-equilibrium steady-states, which is based on the time reversed dynamics \cite{HJ1}. 

\noindent In the present work we shall modify the Onsager-Machlup theory for stationary non-equilibrium states to include cases with  a nonlinear macroscopic equation. We shall generalize the minimum dissipation principle to relaxation processes to steady states that are only locally in equilibrium. The hypothesis of local equilibrium guarantees the existence of the local entropy functional, as well as the reversibility of the microscopic physics. However we do not assume that the transport coefficient $g_{\mu\nu}$'s are independent of the forces $X^\mu$, and so we are not restricted to the linear Onsager region. Our systems will be prevented from reaching equilibrium by being coupled to several reservoirs held at distinct values of the intensive variables, so that the corresponding forces will be non vanishing in our system. These forces will yield fluxes via (\ref{J}), which will transport some of the quantities $\alpha_\mu$ between the reservoirs. The steady-state satisfies the relation satisfies the relation

\begin{equation}
J_\mu{\dot X}^\mu={\dot \alpha}_\mu {\dot X}^\mu=0
\end{equation}
Hence, in a stationary state, some of the currents ${\tilde J}_\mu={\dot{\tilde\alpha}}_\mu\equiv{\dot\beta}_\mu$ will vanish, those orthogonal to the non-vanishing forces using the metric $g_{\mu\nu}$ evaluated at the stationary value of the forces. We shall show that if only this second class of quantities ${\tilde\alpha}_\mu\equiv\beta_\mu$ are perturbed from their stationary state, then their relaxation will extremize $M$ varied over only this second class of currents.  Note that the nonlinear dynamics is nontrivial, because the subspace of the currents which is varied is the orthogonal compliment of the forces between the reservoirs, and this compliment depends on the full force-dependent metric $g_{\mu\nu}$. The extension of the minimum dissipation principle to the nonlinear regime is part of the generalized thermodynamical program reviewed in Ref.~\cite{Giorgio}. In particular, a geometric analogue, known as the minimum rate of dissipation principle, was introduced in Ref.~\cite{us}. A different generalization of the minimum dissipation principle to non-equilibrium steady states has appeared in Ref.~\cite{Romani}.  They continue to work in the linear region, but they relax the local equilibrium condition so that the microscopic dynamics is not necessarily reversible.  This leads to a correction to the flux terms in the quantity $M$, subtracting out the irreversible part.  Instead of basing their proof on a stochastic process, which would be difficult away from local equilibrium, they use the Hamilton-Jacobi equations of Ref.~\cite{HJ1}, \cite{HJ2} which describe the most probable trajectory by which a fluctuation is created.

In Sec.~\ref{generalization} we show that even in the nonlinear case the minimum dissipation principle determines the evolution of the quantities that relax to equilibrium, and then we adapt the Onsager-Machlup derivation of the minimum dissipation principle to the nonlinear case. To avoid misunderstandings, the {\it Minimum Dissipation Principle} herein expressed should be understood as an {\it approximate variational principle in nonlinear transport for systems out of equilibrium} and not as a variational principle, which is rigorously satisfied for systems out of the linear (Onsager) region. A discussion about an estimation of the error can be found in the sub-Sec.~\ref{error}. Then in Sec.~\ref{ex} an example is provided which is a nonlinear version of that of Ref.~\cite{HJ1}.  We shall consider a 2-dimensional, boundary driven, nonlinear zero-range diffusion in which there are only reservoirs on two opposite faces.  We shall see that the minimum dissipation principle is obeyed for diffusion along a direction which depends on the full nonlinear metric.

\vskip0.1truecm
\noindent $\bullet$ {\bf Equilibrium and Boltzmann's Principle}
\vskip0.1truecm
\noindent Consider a closed system which is macroscopically characterized by a vector of extrinsic quantities $\alpha_\mu$.  Motivated by Boltzmann's kinetic theory of gases \cite{boltzmann}, Planck has defined the corresponding entropy $S(\alpha)$ to be, up to a constant shift, proportional to the logarithm of the number of microstates in the microcanonical ensemble which yield the macrostate described by $\alpha_\mu$ \cite{planck}.  In a classical system the number of microstates is infinite, but he suggests that they be coarse grained into discrete quantities which can then be counted.  This has the advantage that it allows him to accurately describe black body radiation, but also that the discretization procedure allows him to define a notion of probability (and in particular a measure) in an apparently deterministic system.   More precisely, he recasts Boltzmann's principle as the ergodicity assumption that, for some choice of discretization, the probability of each allowed microstate will be equal.

\noindent In a very large system, to which one may apply the thermodynamic approximation, there will be a small region in the space of values of $\alpha$, peaked about some value $\alpha^{(0)}_\mu$, for which the entropy is much larger than even the integral over the rest of $\alpha$-space. As $\alpha^{(0)}_\mu$ is a maximum of the entropy function, it satisfies
\begin{equation}
\frac{\partial\Delta S(\alpha)}{\partial \alpha_\mu}\mid_{\alpha_\mu^{(0)}}=0.
\end{equation}
Boltzmann's principle then states that, given no other information about the system, in the thermodynamic limit of a large number of microstates the system will certainly be found in the macrostate $\alpha^{(0)}_\mu$, which is referred to as the equilibrium state.  
\vskip0.1truecm
\noindent $\bullet$ {\bf Relaxation in the Onsager Region}
\vskip0.1truecm
\noindent If on the other hand one begins with an initial condition $\alpha^{(1)}_\mu\neq\alpha^{(0)}_\mu$, at time $t=0$, then the probability that the state is in a given state $\alpha\prime$ at time $t$ is given by a conditional probability, in contrast with the absolute probability in equilibrium. However, due to the ergodicity assumption, if one waits an infinite amount of time then the information about the original state $\alpha^{(1)}_\mu$ is erased and so again the measured state will be the equilibrium state $\alpha^{(0)}_\mu$ with probability $1$. If the conditional probabilities can be calculated, then one can determine not only the final state, but also the full time-dependent trajectory followed by the macrostate from $\alpha^{(1)}_\mu$ to $\alpha^{(0)}_\mu$.  

\noindent How does such a calculation proceed? The tendency of the system to seek equilibrium is measured by the thermodynamic forces $X^\mu$
\begin{equation}
X^\mu(\alpha)=\frac{\partial\Delta_I S(\alpha)}{\partial \alpha_\mu}\mid_{\alpha^{(1)}_\mu}\neq 0.
\end{equation}
where we have introduced the entropy production $\Delta_I S$, and the fact that at $\alpha_\mu=\alpha^{(1)}_\mu$, the system is not at the local equilibrium state. The fluxes (of matter, heat, electricity) are measured by the time derivatives of the $\alpha$'s: $J_\mu(\alpha)=\dot\alpha_\mu$. As $\alpha_\mu$ in general is a vector, so is $X^\mu(\alpha)$. $\dot\alpha_\mu$ depends on $X^\mu$ such that when $X^\mu$ vanishes, so does $\dot\alpha_\mu=J_\mu$.  The essential physical assumption about the irreversible processes is that the fluxes depend on the forces through the transport coefficients $g_{\mu\nu}(X(\alpha),\alpha)$:
\begin{equation}
J_\mu(\alpha)=\frac{d\alpha_\mu}{dt}=g_{\mu\nu}(X(\alpha),\alpha)X^\nu(\alpha).\label{jeq} 
\end{equation}
where, here and in the sequel, the summation convention on the repeated indices is understood. The conditional probabilities can be used to calculate the transport coefficients which then reduce the relaxation of the system to the solution the a system of coupled, nonlinear, first order differential equations (\ref{jeq}). The entropy production is a function of the $\alpha$'s: $\Delta_IS=\Delta_IS(\alpha_1,\cdots,\alpha_n)$. Hence,
\begin{equation}
\frac{d\Delta_IS}{dt}=\frac{\partial \Delta_IS}{\partial\alpha_\mu}\frac{d\alpha_\mu}{dt}=X^\mu J_\mu.
\end{equation}
In general it is difficult or impossible to determine these conditional probabilities, and so this approach is of limited use. However near thermodynamic equilibrium, the entropy production is near its extreme value, and it can be expanded about $\alpha_\mu^{(0)}$
\begin{equation}
\Delta_IS(\alpha)\sim \Delta_IS_0- s^{\mu\nu}\alpha_\mu\alpha_\nu\quad ; \quad
X^\mu(\alpha)=-2s^{\mu\nu}\alpha_\nu \label{quad}
\end{equation}
where $s^{\mu\nu}$ is a constant matrix. In this case, using the fact that $X^\mu(\alpha^{(0)})=0$, one may approximate the current to be
\begin{equation}
\frac{d\alpha_\mu}{dt}=g_{\mu\nu}(X(\alpha),\alpha)X^\nu(\alpha)\sim -2g_{\mu\nu}(0,\alpha^{(0)}) s^{\nu\kappa} \alpha_\kappa.
\end{equation}
This is a system of linear differential equations depending upon two constant matrices.
\vskip0.1truecm
\noindent $\bullet$ {\bf Onsager Machlup Principle}
\vskip0.1truecm
\noindent In Ref. \cite{Onsager} Onsager has generalized a theorem of Lord Rayleigh \cite{Rayleigh}, that relaxation to equilibrium can be derived from a variational principle. They found that  Eq. (\ref{jeq}) can be derived as an extremum of the quantity with respect to the currents $J_\mu$
\begin{equation}
M=X^\mu J_\mu-\frac{1}{2}g^{\mu\nu}J_\mu J_\nu
\end{equation}
where $g^{\mu\nu}$ is the inverse matrix of the transport coefficients. We get
\begin{equation}
0=\frac{\partial M}{\partial J_\mu}=X^\mu-g^{\mu\nu}J_\nu.
\end{equation}
More than 20 years later an interpretation of this formula, near thermodynamic equilibrium, was demonstrated.  In Ref. \cite{OM} Onsager and Machlup provided a stochastic demonstration that close to equilibrium, the probability $f$ of relaxing from $\alpha_1$ to $\alpha_0$ along a path $\alpha(t)$ is proportional to the exponential of $M$ or more precisely
\begin{equation} \label{dist}
f\propto \textup{exp}\left(\frac{1}{2k_B}\int dt\ (X^\mu J_\mu- \frac{1}{2}g^{\mu\nu}J_\mu J_\nu-\frac{1}{2}g_{\mu\nu}X^\mu X^\nu)\right).
\end{equation}
Thus, just as Boltzmann's principle provides a formula for the absolute probability of realizing a certain microstate at a fixed time, Onsager and Machlup found a formula for the probability of following a succession of events during relaxation to equilibrium.

\noindent How did Onsager and Machlup demonstrate that (\ref{dist}) indeed provides the probability of any given sequence of configurations?  Their demonstration rested upon three assumptions.  First, the sequence of events is Markovian. This means that given some set of data at time $t_0$, which in Ref. \cite{OM} is the vector $\alpha$ but the entire approach can be generalized to include their time derivatives in systems with inertia \cite{MO}, the conditional probability of a configuration at any future time $t_1$ is independent of any knowledge of the state at times before $t_0$.  Thus while the theory remains nondeterministic because it the configuration at $t_0$ does not determine that at $t_1$, nonetheless it does not contain any hidden variables at $t_0$ which might affect $t_1$.  In this sense the $\alpha$ provide a complete set of states.  The Markovian property implies that to determine the probability of a trajectory, one only needs the conditional probability of one state given another.

\noindent The second assumption is that the entropy production function is quadratic, as in Eq. (\ref{quad}). This is only a technical assumption to allow a calculation of a path integral in closed form. Given a form of the entropy which is not quadratic, perturbation theory could be applied to the path integral to yield a result as an asymptotic series, as was already known at the time of Onsager and Machlup's paper.  

\noindent Finally they assumed that each extrinsic variable is a sum of local, uncorrelated variables.  As a result, steps in the evolution of these variables, while random, will obey a Gaussian distribution in the thermodynamic limit. This Gaussian distribution is evident in the fact that (\ref{dist}) is expressed as an exponent of squares.

\section{\bf Relaxation to a Steady State}

For more than half a century there have been attempts to generalize Onsager and Machlup's variational principle to the process of relaxation not to equilibrium, but to a stationary state. A stationary state is a state satisfying the condition $J_\mu\delta X^\mu=0$. Hence, at the steady state, not all the thermodynamic forces, $X^\mu$, vanish. Glansdorff and Prigogine have argued that relaxation to a stationary state is not described by an extremization problem in Ref. \cite{GP71}.  Indeed in Ref. \cite{MS} \v{S}ilhav\'{y} states that in a stationary state the thermodynamic quantities themselves are ambiguous.

\noindent The basic problem is the microscopic state is no longer simply an unknown element of the microcanonical ensemble.   Information about the microstate is constantly destroyed by interaction with the outside of the system.  Thus instead of being characterized only by the internal degrees of freedom, the evolution of a microscopic state also depends on external degrees of freedom.  Furthermore some of the information about the internal degrees of freedom cease to be measurable on the inside, as they are transfered to the outside, where no measurements occur. Thus the starting point of this analysis, a definition of the entropy as a sum of the number of microstates, is already ill-defined.

\noindent In this note we will claim that nonetheless a subset of the system's information is approximately determined by an extremization principle.  We will provide a criterion which describes when this approximation is reliable.

\section{\bf Example: A Partially Ionized Plasma}

Consider a closed system containing weakly ionized hydrogen gas in a sealed container with perfectly reflective walls.  The system contains $N_H$ neutral hydrogen atoms, $N_e=N_i$ free electrons and protons as well as a number of free photons, which we will call $N_\gamma$, making the crude approximation that the energy of each photon is equal to the binding energy of an electron in hydrogen.  Similarly we will assert that all of the bound electrons are in a 1s orbital.  Collisions and photon absorption cause the hydrogen to ionize, but also electrons and protons and recombine into a hydrogen atom and a photon.  This photon is just at the right energy to ionize another hydrogen, although usually it needs to bounce off of the perfectly reflecting walls a few times first.  

\noindent The microstates correspond to the discretized positions and velocities of the various particles.   Summing over these positions one can calculate the independent macroscopic variables $N_H$, $N_e$ and $N_\gamma$.  These are components of the vector $\alpha$.   Given some information, like the size of the system and the total energy, one can in principle calculate the number of microstates corresponding to each value of $\alpha$, and so determine the functional form of the entropy.  Maximizing this entropy one can find the state $\alpha_0$ which yields, for example, the ionization fraction of the hydrogen in equilibrium.   It will be approximately given by the Saha equation.

\noindent A short burst with a laser can ionize some atomic hydrogen, taking the system out of the old equilibrium.  A slight expansion of the cavity can remove this additional energy from the system.  Now $N_e$ will be higher than its equilibrium value and $N_i$ will be lower.  One can understand the relaxation of this excited state to the new equilibrium using the analysis of Onsager and Machlup described above.  The excess of free protons and free electrons corresponds to a nonvanishing gradient of the entropy function, which is a thermodynamic force.  This force causes a current, which in this case is just an increase in $N_H$ at the expense of $N_i$.   The total energy is conserved, this change occurs because there are simply more states available with $\alpha$ at its equilibrium level $\alpha_0$ than at the excited level.  

\noindent How can a nonequilibrium steady state be constructed in this example?  Imagine that the laser is permanently turned on, but that the plasma is allowed to transfer some kinetic energy to the walls of the container.  For simplicity we will assume that these collisions are sufficiently elastic that they do not affect the rates of ionization and recombination.  As a result the time derivative of the number of photons $N_\gamma$ will not be determined entirely by the thermodynamic forces, there will also be a contribution from the external current.  However, after waiting for a sufficiently long time, the plasma will heat to a temperature at which $\dot\alpha=0$.

\noindent What can Onsager's formalism tell us about this situation?  Certainly it cannot tell us photon current $\dot{N}_H$, which has an external contribution.  However the the numbers $N_H$ and $N_e$ of hydrogen atoms and free electrons are determined entirely by the internal physics.   This consists of the same equations as in the case of a closed system, as the recombination and ionization processes are unaffected.  All that has changed is the number of photons $N_\gamma$.  But near the steady state, $N_\gamma$ lies close to a known value.

\noindent In particular the number of classical states available at a given moment $S(\alpha)$ is still well-defined.  These are not Hamiltonian eigenstates of a quantum theory which require an infinite amount of time to be manifested, they are coarse grained positions and velocities which exist instantaneously.  The ergodicity assumption means that the interactions, in this case recombination and ionization, will act so as to increase this number of states $S$. 
After all the rate of ionization and recombination are independent of whether the photon arrived from an external source, or was always there bouncing off of the walls.  Thus not only the energy, but also the thermodynamic forces $X$ and the transport coefficients are given by the same formulas as in the case of relaxation to equilibrium.

\noindent The only step in the logical sequence which differs is then the current itself.  The time derivative of $N_\gamma$ has an external contribution.  But the other currents are unaffected, and so continue to be given by Eq. (\ref{jeq}).

\section{The Generalization}\label{generalization}

\subsection{A Functional for the Internal Quantities}

Now what happens if we perturb the system from a stationary state and try to follow Onsager and Machlup's stochastic derivation of the probabilities of various relaxation processes?

\noindent The first problem that arises is that in a stationary state, $\Delta S_I$ is not extremized.  By definition the time derivatives $\dot\alpha_\mu$ vanish at a stationary state.  However this no longer implies that the forces vanish. We can project the vector $\alpha_\mu$ into two vectors $\beta_\mu$ and $\gamma_\mu$.  Here $\gamma_\mu$ represents the quantities which are subject to external sources, and $\beta_\mu$, as in a closed system, are absolutely conserved. In the plasma example, $\beta_\mu$ contains $N_h$ and $N_e$ and $\gamma_\mu$ contains $N_\gamma$.  It is not difficult to show that for any given configuration $\alpha_\mu$ we can choose a basis of the $\gamma_\mu$ orthogonal to all of the $\beta_\mu$ with respect to the transport coefficients $g_{\mu\nu}$. Indeed, the evolution equation for fluctuations $\alpha_\mu$ may be brought into the form
\begin{equation}\label{fq1}
\frac{d\alpha_\mu}{dt}=g_{\mu\nu}\frac{\partial\Delta_I S}{\partial \alpha_\nu}+J_\nu^{(ext.)}=g_{\mu\nu}\frac{\partial\Delta_I S}{\partial \alpha_\nu}+g_{\mu\kappa}M^\kappa_\nu X^{\nu \rm{(ext)}}
\end{equation}
\noindent with $J_\nu^{(ext.)}\equiv g_{\nu\kappa}M^\kappa_\eta X^{\eta(ext.)}$. The greek indexes $\mu,\ \nu,\ \kappa,\ \cdots$ run from $1,\cdots , n$, and the second term on the right-hand side of Eq.~(\ref{fq1}), $J_\nu^{(ext.)}$, takes into account the contribution to the dynamics due to thermodynamic forces held fixed by the boundary conditions. Note that the first $m$ components of the column vector $X^{\nu{\rm{(ext)}}}$ correspond to the $m$ (with $m<n$) thermodynamic forces held fixed by the boundary conditions, whereas the remaining $n-m$ components of this vector are set to zero. The vector $\alpha_\mu$ may be projected into two fluctuating vectors $\beta_\mu$ and $\gamma_\mu$ by introducing two matrices $A^\mu_\nu$ and $B^\mu_\nu$, which are orthogonal with respect to the metric defined by the transport coefficients. In particular, we define
\begin{equation}\label{fq2}
\beta_\mu\equiv A^\nu_\mu\alpha_\nu\qquad ; \qquad\gamma_\mu\equiv B^\nu_\mu\alpha_\nu
\end{equation}
\noindent with matrices $A^\mu_\nu$ and $B^\mu_\nu$ satisfying the orthogonal conditions
\begin{equation}\label{fq3}
A^{\kappa}_\mu g_{\kappa\eta}B^\eta_\nu=0\quad{\rm and}\quad A^\kappa_\mu g_{\kappa\eta}M^\eta_\nu=0
\end{equation}
\noindent The matrices $A^\mu_\nu$ and $B^\mu_\nu$ are related to their inverse matrices through the relations 
\begin{equation}\label{fq4}
g^{\mu\eta}A^\kappa_\eta g_{\kappa\tau}A^\tau_\nu=A^{\mu\kappa}A_{\kappa\nu}=\delta^\mu_\nu\qquad{\rm and}\qquad g^{\mu\eta}B^\kappa_\eta g_{\kappa\tau}B^\tau_\nu=B^{\mu\kappa}B_{\kappa\nu}=\delta^\mu_\nu
\end{equation}
\noindent with $\delta^\mu_\nu$ denoting Kronecker's delta and the metric tensor $g_{\mu\nu}$ is used for raising and/or lowering the indexes. Note that from Eq.~(\ref{fq4}) we get the useful relations
\begin{equation}\label{fq5}
A^\kappa_\mu g_{\kappa\eta}A^\eta_\nu=g_{\mu\nu}\quad{\rm and}\quad B^\kappa_\mu g_{\kappa\eta}B^\eta_\nu=g_{\mu\nu}
\end{equation}
\noindent By taking into account Eqs~(\ref{fq2}) and (\ref{fq5}), the symmetry relation $g_{\mu\nu}=g_{\nu\mu}$, and the identity
\begin{equation}\label{fq6}
\frac{\partial \Delta_I S}{\partial\alpha_\mu}=A^\mu_\nu\frac{\partial\Delta_I S}{\partial \beta_\nu}+B^\mu_\nu\frac{\partial\Delta_I S}{\partial \gamma_\nu}
\end{equation}
\noindent it is easily checked that in the basis $\beta_\mu$ and $\gamma_\mu$, the evolution equations (\ref{fq1}) factorize
\begin{equation}
\frac{d\beta_\mu}{dt}=g_{\mu\nu}\frac{\partial\Delta_I S}{\partial \beta_\nu}\qquad ; \qquad
\frac{d\gamma_\mu}{dt}=g_{\mu\nu}\frac{\partial\Delta_I S}{\partial \gamma_\nu}+B^\nu_\mu J_\nu^{(ext.)}
\end{equation}
\noindent As the evolution of $\gamma_\mu$ is affected by external forces, it cannot be determined even statistically by a thermodynamic argument of the system itself \cite{GP71}. The most naive approach would be to simply apply Onsager's formalism to $\beta_\mu$, ignoring the $\gamma_\mu$. After all the evolution of $\beta_\mu$ in a relaxation to a steady state is described by equations which look like those describing the evolution of $\alpha_\mu$ in a relaxation to equilibrium.

\noindent Recall that Onsager and Machlup's derivation rested upon three pillars. The third, that the extrinsic variables are sums of uncorrelated variables, is unaffected. The definitions of these variables do not depend upon the external forces.   

\noindent However there is one critical difference between the equations describing $\alpha_\mu$ and $\beta_\mu$. In the later case, the entropy production  $\Delta_I S(\beta_\mu,\gamma_\mu)$ is a function not just of $\beta_\mu$, but also of $\gamma_\mu$. This is a problem for the first two pillars. The $\beta_\mu$ are no longer a complete set, and so the first pillar, the Markovian property featured in Onsager and Machlup's derivation is lost. The $\gamma_\mu$ are hidden variables and they create a systematic bias not only in the statistical fluctuations, but also in the mean evolution of the $\beta_\mu$.  What about the second pillar? The entropy still extremizes $\beta_\mu$, and in a stationary solution $\gamma_\mu$ is fixed to some value $\gamma_\mu^0$.  Setting $\gamma_\mu=\gamma_\mu^0$, the action may still be expanded in $\beta_\mu$ and it will be quadratic. However $\gamma$ will change during the relaxation process, and likely during the excitation away from the steady state. Thus there will be corrections to the entropy which depend upon $\gamma_\mu-\gamma_\mu^0$.

\noindent Both of these problems have the same cause. The evolution of $\gamma_\mu$ during the relaxation is unknown.  It depends not only on the internal physics, but also on the external source.  As the entropy couples $\beta_\mu$ and $\gamma_\mu$, this causes a finite error in an estimate of the trajectory of the $\beta_\mu$'s alone using Onsager and Machlup's functional.

\subsection{Estimating the Error}\label{error}

Just how big is this error?  When must it be considered?  Of course it depends on just how much $\gamma$ deviates from its stationary value $\gamma_0$, which depends on a physical choice.  We are interested in a system which is excited from a steady state and then decays.  In the case of our partially ionized plasma subjected to a constant laser, for example the laser may be turned off for a moment, taking the system out of its steady state.  There will momentarily be some additional recombination and the temperature will drop.  But eventually, as less kinetic energy will be transferred to the walls, the system will relax to its steady state.

\noindent In this case the dependence upon the unknown physics of the external photon injection is essential.  The perturbation from the steady state itself changes the photon number $N_\gamma$, which is $\gamma$ in this example.  It is essential to know just how $\gamma$ returns to $\gamma_0$.  This restoration depends strongly on the physics of the photon injection into the system.  It cannot, to any approximation, be described by the internal physics of the system. Therefore, an application of the Onsager Machlup to the evolution in this case appears to be at best difficult, as the information from the functional needs to be supplemented by information about the laser.

\noindent On the other hand, one can also consider a perturbation from a steady state in which one only changes $\beta$, leaving $\gamma=\gamma_0$.  If $\gamma$ continues to be equal to $\gamma_0$ during the entire relaxation process, then the Onsager Machlup functional applied to $\beta$ will function as well as in a relaxation to equilibrium.  Of course $\gamma$ will not be equal to $\gamma_0$ throughout the relaxation, even if it is equal at the beginning and the end. {\it{It is this transitory deviation of $\gamma$ from $\gamma_0$ which contributes an unavoidable error to the Onsager Machlup functional for the probability of a given relaxation to a steady state.}}

\noindent We can estimate the error by expanding the entropy to leading order about the steady state values of $\beta$ and $\gamma$, which for simplicity we will set to $\beta_0=\gamma_0=0$. The entropy production is
\begin{equation}
\Delta_I S=\beta a\beta+\gamma b\beta+\gamma c\gamma. \label{seq}
\end{equation}
Now we will perturb $\beta$ to $\beta_1$.  This will exert a force
\begin{equation}
X=\frac{\partial\Delta_I S}{\partial \gamma}= b\beta_1 + 2 c\gamma.
\end{equation}
The force vanishes when
\begin{equation}
\gamma=-\frac{1}{2}c^{-1}b\beta_1. \label{gamma}
\end{equation}
However $\dot\gamma$ is also effected by an external force, which compensates for this effect and so generically can cause an appreciable shift in $\gamma$ from the value in Eq. (\ref{gamma}).  However, while this shift can easily be smaller than or comparable to the value in (\ref{gamma}) itself, without fine tuning it will not be much greater. Therefore Eq. (\ref{gamma}) provides a rough upper limit on the $\gamma$ which can be expected during the relaxation. This will be sufficient for our estimation.

\noindent Now that during the relaxation $\gamma$ will shift, so will the action for $\beta$ itself.  This backreaction of $\gamma$ upon $\beta$ is the obstruction to a variational principle describing the relaxation to a steady state.  To estimate it, we can simply insert Eq. (\ref{gamma}) into Eq. (\ref{seq})
\begin{equation}
\Delta_I S=\Delta_I S^b+\delta\Delta_I S\quad ; \quad \Delta_I S^b=\beta a\beta\quad ; \quad
\delta\Delta_I S=\beta b c^{-1} b \beta.
\end{equation}
Here $\Delta_I S^b$ is the contribution which, using the Stochastic derivation of Ref. \cite{OM}, would lead a probability for each relaxation.  The term $\delta\Delta_I S$ is the correction due to the coupling with $\gamma$.  The correction is comparable to the unperturbed effect when $ac\sim b^2$.   Recall that $a$ and $c$ are the diagonal transport coefficients and $b$ are the off-diagonal coefficients, relating $\beta$ and $\gamma$.  Therefore this is a condition on how close the transport coefficients are to a block diagonal form. herefore, we have learned that

\noindent {\it{The Onsager-Machlup functional, restricted to the quantities $\beta$ which have no external interactions, approximates the probability of a particular relation to a stationary state up to corrections which are suppressed by the square of those transport coefficients which mix the quantities with and without external interactions.}}

\section{Example: Nonlinear Diffusion} \label{ex}

\begin{figure*}[htb] 
\includegraphics[width=13cm,height=8cm]{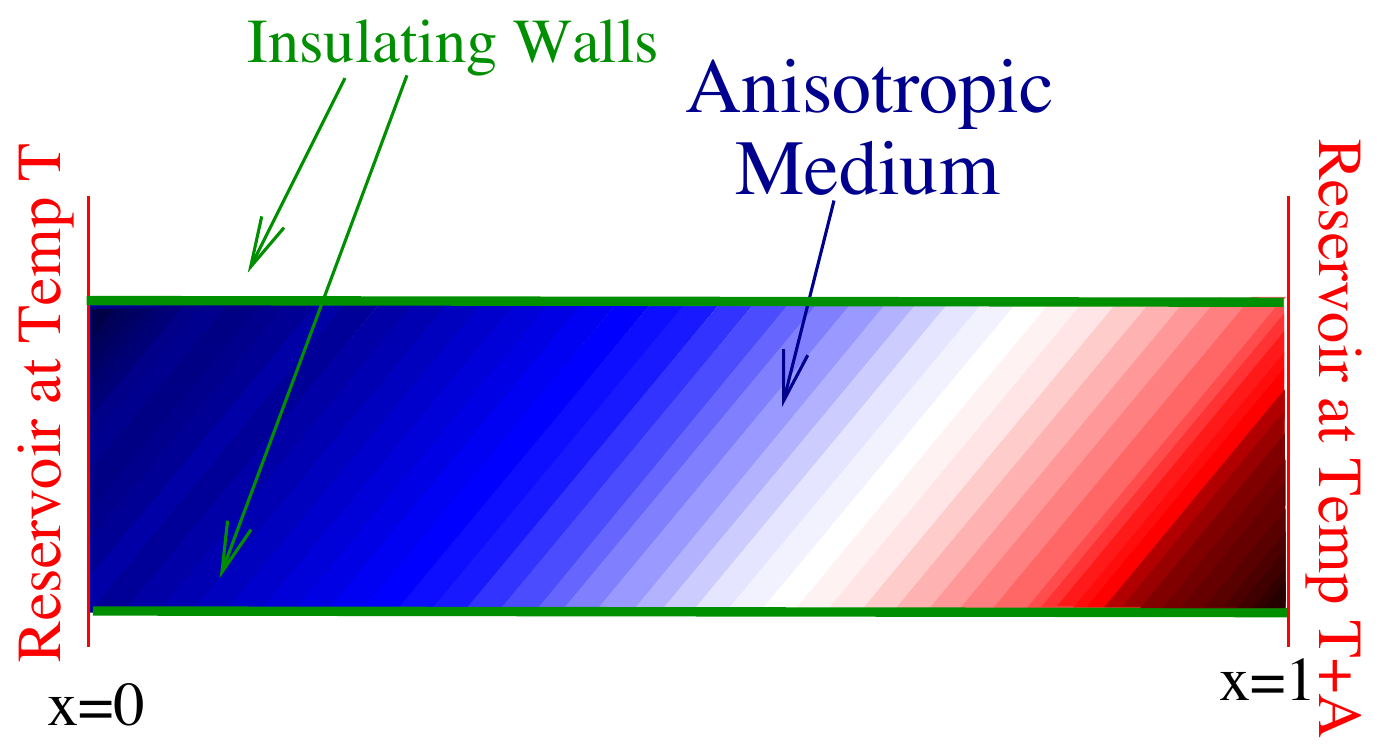}
\caption{\label{diff2}
A rectangular anisotropic medium is placed between two reservoirs at different but similar temperatures.  The other two walls are insulating.  Heat flows between the two reservoirs, establishing a temperature gradient in the $x$ direction.  The transport coefficients are nondiagonal in the $x-y$ basis, and so there is also a gradient in the $y$ direction.  Due to the nonlinearity of the flux-force relation in this example, the angle of the isothermal lines depends on the difference between the temperatures of the reservoirs.}
\end{figure*}

As an example of the minimum dissipation principle at work, consider diffusion in an anisotropic box extending from $x=0$ to $x=L$ in the $x$ direction and in some small interval along the $y$ direction.  On both boundaries of the $x$ coordinate, couple the system to a reservoir. Let the temperature of the reservoir on the right be $A$ units hotter than that of the reservoir on the left. As a result there will be a flux of heat from right to left across our anisotropic medium.  For simplicity, let $A$ be much smaller than either of the two temperatures. On the $y$ boundaries place an insulator. This configuration is depicted in FIG.~\ref{diff2}.

\noindent This two-dimensional heat conduction problem involves two thermodynamic forces, the gradients
\begin{equation}
X_x=\frac{\partial}{\partial x}\frac{1}{T}\quad X_y=\frac{\partial}{\partial y}\frac{1}{T}
\end{equation}
and two currents, which are the heat fluxes $J_x$ and $J_y$ in the $x$ and $y$ directions.  Consider the asymmetric nonlinear transport equations
\begin{equation}
J_x=\lambda_{xx}T^2X_x\quad
J_y=\lambda_{yx}LT^3X_x^2+\lambda_{yy}T^2X_y.
\end{equation}
\noindent Hence, the transport coefficients read
\begin{equation}
g_{xx}=\lambda_{xx}T^2,\qquad g_{xy}=0\quad
g_{yx}=\lambda_{yx}LT^3X_x,\qquad g_{yy}=\lambda_{yy}T^2
\end{equation}
\noindent 
Here the nonlinearity comes from the off-diagonal $\lambda_{yx}$ term, which creates a temperature flux along the $y$ direction as heat flows between the reservoirs.  Let $\lambda_{yx}$ be large, so that the nonlinear effects will not be drown out at big $T$. As the temperature difference between the reservoirs is small with respect to the temperature, there will be a stationary state when the temperature gradient in the $x$ direction is constant
\begin{equation}
X_x=-\frac{A}{LT^2}.
\end{equation}
Now the heat flux in the $y$ direction is
\begin{equation}
J_y=\frac{A^2\lambda_{yx}}{LT}+\lambda_{yy}T^2X_y.
\end{equation}
As we are interested in boundary conditions in which no heat can escape from the boundaries in the $y$ directions, one may impose $J_y=0$ and so
\begin{equation}
X_y=-\frac{A^2\lambda_{yx}}{LT^3\lambda_{yy}}=-\frac{1}{T^2}\frac{\partial T}{\partial y} 
\end{equation}
and so the temperature gradient along the $y$ direction is
\begin{equation}
\frac{\partial T}{\partial y}=\frac{A^2\lambda_{yx}}{LT\lambda_{yy}}.
\end{equation}
As there is no current along the $y$ direction, and as the current in the $x$ direction is constant, the divergence of the current vanishes and so this is a steady state.  Explicitly one finds
\begin{equation}
\frac{\partial T}{\partial t}=-\nabla\cdot {\bf J}=\frac{\partial}{\partial x}\Bigl(\frac{A\lambda_{xx}}{L}\Bigr)-\frac{\partial}{\partial y}0=0
\end{equation}
\noindent While the forces in both the $x$ and $y$ directions are non-vanishing at the steady state, the force in the direction
\begin{equation}
\hat{y}=Ty-A\frac{\lambda_{yx}}{\lambda_{yy}}x
\end{equation}
\noindent is zero. The temperature in this direction is roughly constant.  Notice however, that unlike the linear situation, the direction $\hat{y}$ itself depends on the boundary conditions $A$.  In this case we can factorize the problem into two one-dimensional problems, one in the $x$ direction and one in the $\hat{y}$ direction.  The $\hat{y}$ problem is just that of Onsager and Machlup, but to the aforementioned effects that come from interactions with the other system if one perturbs too far away from the stationary state.  Therefore, for sufficiently small perturbations of the $\hat{y}$ profile, the one-dimensional relaxation obeys the minimum dissipation principle.  More precisely, the relaxation of a perturbation which depends only on $\hat{y}$ extremizes $M$ varied with respect to temperature fluxes along $\hat{y}$. In this case we were very fortunate, because the two-dimensional problem globally factorized into two one-dimensional problems.  This is because the integral curves of the $\hat{y}$ vector form a linear subspace as $\hat{y}$ is the same direction everywhere.  In general, the direction $\hat{y}$ will depend on the $x$ coordinate.  In this case one may still factorize problem, but the integral curves whose tangent vectors are $\hat{y}$ will no longer be straight lines as the $\hat{y}$ direction will be position-dependent.  Thus this factorization only applies locally, but this is sufficient as we have assumed local equilibrium.  Things become even more complicated above two dimensions, if the number of directions along which one varies is at least equal to two.  In this case the vanishing fluxes define a set of vectors at each point, but this set of vectors does not necessarily form the tangent space to any foliation of our space and so, if some integrability condition is not satisfied, the problem cannot be globally factorized into a near-equilibrium problem and a constant problem.  However locally this factorization is always possible and the principle of minimal dissipation follows.

\section{Conclusion}

We have generalized Onsager's minimum dissipation principle to relaxations to steady states which are only locally in equilibrium.  We have used Onsager and Machlup's stochastic method to demonstrate this generalization of the principle, but we were only able to demonstrate it for variations with respect to a subset of the variables for which fluxes are not driven by the boundary conditions. However it appears that the full reciprocal relations are consistent with the vanishing of variations with respect to all of the fluxes. It should be noted that we have not determined quantitatively to which order in the size of the variation the principle holds. 

\noindent  Our generalization differs from that of Ref.~\cite{Romani}, in that we allow nonlinear flux-force relations while they allow microscopic irreversibility. It would be interesting to see if one may formulate a generalization which incorporates both theories. This is impeded by the fact that our stochastic approach is difficult to generalize to the irreversible case, while their Hamilton-Jacobi approach assumes linearity in a number of places, such as the constancy of their diffusion matrices and their quadratic ansatz for the Lagrangian density. The next step is to incorporate this principle into the geometrical thermodynamic field theory of Ref.~\cite{Giorgio}. To do this, it would be useful to relate $M$ to the length of the system's trajectory in some space, so that the minimum dissipation principle becomes the shortest path.

\section*{Acknowledgments}

\noindent One of us (G.S.), is very grateful to Prof. Pasquale Nardone of the Universit{\'e} Libre de Bruxelles for his scientific suggestions. J.E. is supported by NSFC MianShang grant 11375201.

\bibliographystyle{unsrt}

\begin{thebibliography}{alpha}

\bibitem{bertini} L. Bertini, A. De Sole, D. Gabrielli, G. Jona-Lasinio, C. Landim, {\it J. Stat. Phys.}, {\bf 123}, 237 (2006).
\bibitem{appert} C. Appert-Rolland, B. Derrida, V. Lecompte, F. van Wijlandn, {\it Phys. Rev. E}, {\bf 78}, 021122 (2008).
\bibitem{bertini1} L. Bertini, A. De Sole, D. Gabrielli, G. Jona-Lasinio, C. Landim, {\it J. Stat. Mech. Theory Exp.}, {\bf 2007}, P07014 (2007).
\bibitem{bodinau} T. Bodineau, B. Derrida, {\it Phys. Rev. E}, {\bf72}, 066110 (2005).
\bibitem{derrida} B. Derrida, {\it J. Stat. Mech. Theory Exp.}, {\bf 2007}, P07023 (2007).
\bibitem{lanford} O.E. Lanford, ed. Lecture Notes in Physics, {\bf 20} (Springer, Berlin 1973).
\bibitem{kipnis} C. Kipnis, S. Olla and S.R.S. Varadhan, {\it Comm. Pure Appl. Math.}, {\bf 42}, 115 (1989).
\bibitem{jona} G. Jona-Lasinio, C. Landim and M.E. Vares, {\it Prob. Theory Related Fields}, {\bf 97}, 339 (1993).
\bibitem{HJ1} L. Bertini, A. De Sole, D. Gabrielli, G. Jona-Lasinio and C. Landim, {\it J. Stat. Phys.}, {\bf 107} (2002) 635, cond-mat/0108040.
\bibitem{graham} R. Graham and T. T{\'e}l, {\it Phys. Rev. A}, {\bf 33}, 1322 (1986).
\bibitem{eyink} G. Eyink, {\it J. Stat. Phys.}, {\it 61}, 533 (1990).
\bibitem{luchinsky} D.G. Luchinsky and P.V.E. McClintock, {\it Nature}, {\bf 389}, 463 (1997).
\bibitem{prigogine} I. Prigogine, 1954 {\it Thermodynamics of Irreversible processes}, (John Wiley \& Sons).
\bibitem{boltzmann} L. Boltzmann, {\it Sitzungsber. d. k. Akad. d. Wissensch. su Wien} {\bf{II}}76 (1877) 428
\bibitem{landau} L. Landau and E. Lifshitz, {\it Course of Theoretical Physics}, Statistical Physics, {\bf 5} (Pergamon Press, New York, 1968).
\bibitem{degroot} S.R. De Groot and P. Mazur, 1984 {\it Non-Equilibrium Thermodynamics}, Dover Publications, Inc., New York. 
\bibitem{GP71} P. Glansdorff and I. Prigogine, 1971 {\it Thermodynamic Theory of Structure, Stability and Fluctuations}, (Wiley-Interscience). 
\bibitem{Onsager} L. Onsager, {\it Phys. Rev.}, {\bf 37} (1931) 405.
\bibitem{Rayleigh} Lord Rayleigh, {\it Phil. Mag.}, {\bf 26} (1913) 776.
\bibitem{OM} L. Onsager and S. Machlup, {\it Phys. Rev.}, {\bf{91}} (1953) 1505.
\bibitem{MO} S. Machlup and  L. Onsager, {\it Phys. Rev.}, {\bf{91}} (1953) 1512.
\bibitem{Giorgio} G. Sonnino, {\it Phys.Rev. E}, {\bf 79}, 051126, (2009).
\bibitem{us} G. Sonnino and J. Evslin, {\it Int. J. Mod. Chem.}, {\bf 107} (2006) 968.

\noindent G. Sonnino and J. Evslin, {\it Phys. Lett.}, {\bf A365} (2007) 364.
\bibitem{Romani} L. Bertini, A. De Sole, D. Gabrielli, G. Jona-Lasinio and C. Landim {\it Minimum dissipation principle in stationary nonequilibrium states}, cond-mat/0310072.
\bibitem{HJ2} L. Bertini L., A. De Sole, D. Gabrielli, G. Jona-Lasinio and C. Landim, {\it Phys. Rev. Lett.}, {\bf{87}} (2001) 040601, cond-mat/0104153.
\bibitem{planck} M. Planck,  {\it Ueber das Gesetz der Energieverteilung im Normalspectrum}, Annalen der Physik Vol 309 Issue 3 (1901) 553-563.

\noindent M. Planck,  {\it  Annalen der Physik } Vol {\bf 309} Issue {\bf 3} (1901) 553.
\bibitem{MS} M. \v{S}ilhav\'{y}, {\it Arch Ration Mech Anal}, {\bf 68} no. 4 (1978) 299

\end{thebibliography}

\bigskip
\end{document}